\documentstyle[twoside,11pt]{article}
\normalsize
\begin{document}
\bibliographystyle{plain}

\def\greaterthansquiggle{\raise.3ex\hbox{$>$\kern-.75em\lower1ex\hbox{$\sim$}}}
\def\lessthansquiggle{\raise.3ex\hbox{$<$\kern-.75em\lower1ex\hbox{$\sim$}}}
\newcommand{\bfi}{\begin{figure}}
\newcommand{\efi}{\end{figure}}
\newcommand{\npb}{\nopagebreak[3]}
\newcommand{\beq}{\begin{equation}}
\newcommand{\eeq}{\end{equation}}
\newcommand{\bdi}{\begin{displaymath}}
\newcommand{\edi}{\end{displaymath}}
\newcommand{\gaa}{\gamma_{\alpha}}
\newcommand{\gab}{\gamma_{\beta}}
\newcommand{\gam}{\gamma_{\mu}}
\newcommand{\gan}{\gamma_{\nu}}
\newcommand{\gaM}{\gamma^{\mu}}
\newcommand{\gaN}{\gamma^{\nu}}
\newcommand{\gaf}{\gamma_{5}}
\newcommand{\tr}{\mbox{tr}}
\newcommand{\Tr}{\mbox{\it\bf Tr}}
\newcommand{\im}{\mbox{Im}}
\newcommand{\intm}{\int_{4m^2}^{\infty}}
\newcommand{\fpi}{\frac{1}{\pi}}
\newcommand{\range}{\mbox{range}}
\newcommand{\coker}{\mbox{coker}}
\newcommand{\ch}{\mbox{ch}}
\newcommand{\beqa}{\begin{eqnarray}}
\newcommand{\eeqa}{\end{eqnarray}}
\newcommand{\no}{\nonumber}
\newcommand{\bob}{\hspace{0.2em}\rule{0.5em}{0.06em}\rule{0.06em}{0.5em}\hspace{0.2em}}
\newcommand{\grts}{\greaterthansquiggle}
\newcommand{\lets}{\lessthansquiggle}
\def\dddot{\raisebox{1.2ex}{$\textstyle .\hspace{-.12ex}.\hspace{-.12ex}.$}\hspace{-1.5ex}}
\def\Dddot{\raisebox{1.8ex}{$\textstyle .\hspace{-.12ex}.\hspace{-.12ex}.$}\hspace{-1.8ex}}
\newcommand{\Un}{\underline}
\newcommand{\ol}{\overline}
\newcommand{\ra}{\rightarrow}
\newcommand{\Ra}{\Rightarrow}
\newcommand{\ve}{\varepsilon}
\newcommand{\vp}{\varphi}
\newcommand{\vt}{\vartheta}
\newcommand{\dg}{\dagger}
\newcommand{\wt}{\widetilde}
\newcommand{\wh}{\widehat}
\newcommand{\br}{\breve}
\newcommand{\A}{{\cal A}}
\newcommand{\B}{{\cal B}}
\newcommand{\C}{{\cal C}}
\newcommand{\D}{{\cal D}}
\newcommand{\F}{{\cal F}}
\newcommand{\G}{{\cal G}}
\newcommand{\Ha}{{\cal H}}
\newcommand{\K}{{\cal K}}
\newcommand{\cL}{{\cal L}}
\newcommand{\M}{{\cal M}}
\newcommand{\R}{{\cal R}}
\newcommand{\dfrac}{\displaystyle \frac}
\newcommand{\hy}{${\cal H}\! \! \! \! \circ $}
\newcommand{\h}[2]{#1\dotfill\ #2\\}
\newcommand{\tab}[3]{\parbox{2cm}{#1} #2 \dotfill\ #3\\}
\def\au{{\setbox0=\hbox{\lower1.36775ex%
\hbox{''}\kern-.05em}\dp0=.36775ex\hskip0pt\box0}}
\def\ao{{}\kern-.10em\hbox{``}}
\def\lint{\int\limits}

%
%
%
%
%
\newcommand{\dsla}{\partial\hspace{-6pt} /  }  
\newcommand{\Asla}{A\hspace{-6.5pt}  /  } 
\newcommand{\Dsla}{D\hspace{-7.3pt}  /  } 
\newcommand{\Qsla}{Q\hspace{-7.2pt}  /  } 
\newcommand{\psla}{p\hspace{-5.375pt} /   } 
\newcommand{\ksla}{k\hspace{-6pt} /  }
\newcommand{\qsla}{q\hspace{-6pt} /   } 
\newcommand{\asla}{a\hspace{-6.25pt} /   } 
\newcommand{\rsla}{r\hspace{-5pt} /   } 
\newcommand{\ssla}{s\hspace{-5pt} /   } 
\newcommand{\lsla}{l\hspace{-4.5pt} /   } 
\newcommand{\tsla}{t\hspace{-4.75pt} /   } 
\newcommand{\eps}{\epsilon\hspace{-5.325pt} / }
\newcommand{\zsla}{z\hspace{-6pt} /  }
\newcommand{\DDsla}{D\hspace{-5.83pt}  /  }
\newcommand{\ddsla}{\partial\hspace{-4.6pt} /  }  
\newcommand{\DDsq}{D\hspace{-5.83pt}  / \hspace{2.2pt} ^2 }
\newcommand{\Dsq}{D\hspace{-7.3pt}  /  \hspace{2.5pt} ^2 } 
\newcommand{\AAsla}{A\hspace{-5pt}  /  }
\newcommand{\QQsla}{Q\hspace{-5.8pt}  /  }

\begin{titlepage}
\begin{flushright} 
BUTP-95/25
\end{flushright}
\begin{center}

{\Large Perturbative solution of the Schwinger model}\\

\bigskip

Christoph Adam \\
Institut f\"ur theoretische Physik, Universit\"at Bern \\
Sidlerstra\ss e 5, CH-3012 Bern, Switzerland$^*)$ \\

\bigskip

\medskip

\bigskip

{\bf Abstract} \\

\bigskip

For the exactly solvable Schwinger model one interesting 
question is how to infer the exact solution from perturbation theory.
We give a systematic procedure of deriving the exact solution from Feynman
diagrams of arbitrary order for arbitrary $n$-point functions. As a
byproduct, from perturbation theory we derive exact integral equations 
that the $n$-point functions have to obey.

\vfill

\end{center}
$^*)${\footnotesize permanent address: Institut f\"ur theoretische Physik,
Universit\"at Wien \\
Boltzmanngasse 5, 1090 Wien, Austria \\
email address: adam@pap.univie.ac.at}
\end{titlepage}

\section{Introduction}

The Schwinger model \cite{Sc1}, which is QED$_2$ with one massless fermion, has devoted a
long history of extensive study. The reason for this is that the Schwinger 
model combines two features that make it an ideal test labor for field theory
studies and phenomenological methods. On one hand, it is exactly solvable and
solutions within the operator formalism \cite{LS1} or the path integral
framework (\cite{Jay}, \cite{SW1}, \cite{Adam}, \cite{Diss}, \cite{DSEQ}) are
wellknown. On the other hand, it shares some nontrivial features with more
realistic field theories: An anomaly is present and may be used for
quantization of the fermion (\cite{GS1}, \cite{Le1}, \cite{ABH}, \cite{DR1},
\cite{Diss}). Further instantons are present (\cite{Jay}, \cite{SW1},
\cite{Sm1}, \cite{Adam}, \cite{Diss}) and cause the formation of the fermion
condensate and of a nontrivial vacuum ($\Theta$-vacuum) (\cite{LS1},
\cite{Jay}, \cite{SW1}, \cite{Sm1}, \cite{Adam}, \cite{Diss}, \cite{DSEQ}). 
Besides, confinement of the fermion is realized in a precise manner: the
fermionic $n$-point functions do not tend to free ones for large separations
(\cite{LS1}, \cite{KS1}, \cite{CKS}, \cite{KS2}). 

One interesting question in this context is whether or how far it is possible
to derive the exact solution and its nontrivial -- partly nonperturbative --
features from perturbation theory. 

This question we investigate, and we find that by a systematic procedure we
may transform all graphs of a given order into the corresponding contribution
to the exact $n$-point function at hand. Further, from an investigation of
Feynman graphs of arbitrary order we prove some exact integral equations to
hold for specific $n$-point functions. These are related to the
Dyson-Schwinger equations of the Schwinger model (see e.g. \cite{DSEQ}).  

The organization of the paper is as follows: first we review the exact path
integral solution of the Schwinger model in Euclidean space and show how even
contributions from nontrivial vacuum sectors may be derived from the trivial
sector, being thereby accessible to perturbation theory.

Then we present the perturbative approach to the Schwinger model and derive
the rules of how to reduce Feynman graphs to the exact solution. These
computations turn out especially simple in a graphical fashion.

At last, from perturbation theory, we derive some exact integral equations for
the $n$-point functions and comment on their meaning.

\section{The exact path integral solution}

As a starting point we use the exact vacuum functional after integration of
the fermions (which can be found e.g. in \cite{Adam}, \cite{DSEQ}),
\begin{displaymath}
Z[\lambda ,\eta ,\bar\eta ]=\sum_{k=-\infty}^\infty Z_k [\lambda ,\eta
,\bar\eta ],
\end{displaymath}
\begin{displaymath}
Z_k [\lambda ,\eta ,\bar\eta ]=N\int D\beta^{[k]}\prod_{i_0 =0}^{k-1}
(\bar\eta \Psi_{i_0}^{[\beta]})(\bar\Psi_{i_0}^{[\beta]}\eta)\cdot
\end{displaymath}
\beq
\cdot e^{\int dxdy\bar\eta (x)G^{[\beta]}(x,y)\eta (y)} e^{\int dx(\frac{1}{2}
\beta {\rm\bf D}\beta +\frac{1}{e}\beta\lambda)}
\eeq
where the gauge field is parametrized by the prepotential $\beta$
\begin{displaymath}
A_\mu =\frac{1}{e}\epsilon_{\mu\nu}\partial^\nu \beta \quad ,\quad
A_\mu J^\mu =:\beta\lambda
\end{displaymath}
\beq
\Rightarrow\frac{\delta}{\delta J_\mu}=\epsilon^{\mu\nu}\partial_\nu
\frac{\delta}{\delta\lambda}
\eeq
corresponding to Lorentz gauge. $G^{[\beta]}$ is the exact fermion propagator
\begin{displaymath}
G^{[\beta]}(x,y)=e^{i(\beta(x)-\beta(y))\gaf}G_0 (x-y),
\end{displaymath}
\beq
G_0 (z)=\frac{z^\mu \gam}{2\pi (z^2 -i\epsilon)}.
\eeq
{\bf D} is the operator of the effective action of the prepotential $\beta$,
with Green's function {\bf G},
\begin{displaymath}
{\rm\bf D}=\frac{1}{\pi \mu^2}\Box (\Box -\mu^2)\quad ,\quad
\mu^2=\frac{e^2}{\pi} 
\end{displaymath}
\begin{displaymath}
{\rm\bf G}(x)=\pi (D_\mu (x)-D_0 (x))
\end{displaymath}
\beq
{\rm\bf D}_x {\rm\bf G}(x-y)=\delta (x-y)
\eeq
where $\mu$ is the "photon" mass and $D_\mu (x),D_0 (x)$ are the massive and
massless scalar propagators in two dimensions:
\beq
D_0 (x)=\frac{1}{4\pi}\ln (x^2 -i\epsilon) +\mbox{const. }\quad ,\quad
D_\mu (x)=-\frac{1}{2\pi}K_0 (\mu |x|),
\eeq
$K_0 (z)$ being the McDonald function (for details see e.g. \cite{Adam},
\cite{Diss}, \cite{DSEQ}).

$k$ is the instanton number and $\Psi_{i_0}^{[\beta]}$ are the zero modes
corresponding to the gauge field $\beta$ with instanton number $k$. (Their
precise form can be found in \cite{Adam}, \cite{Diss}, \cite{DSEQ}).

Due to the Gaussian nature of the path integral (1) and the knowledge of the
exact fermion propagator in an external field (3) all $n$-point functions of
the theory may be computed explicitly. E.g. for the fermionic two-point
function we find (for the sectors $k=0,\pm 1$; higher sectors do not
contribute) 
\beq
\langle T\bar\Psi^\beta (y)\Psi^\alpha (x)\rangle^{k=0}=G_0^{\alpha\beta}(y-x)
e^{{\rm\bf G}(0)-{\rm\bf G}(x-y)}
\eeq
\beq
\langle T\bar\Psi^\beta (y)\Psi^\alpha (x)\rangle^{k=\pm 1}=\frac{1}{2\pi}
P_\pm^{\alpha\beta}e^{{\rm\bf G}(0)+{\rm\bf G}(x-y)}.
\eeq

We observe a feature that persists to hold for higher $n$-point functions: for
vector-like fermionic bilinears only the zero sector contributes, whereas for
chiral bilinears the contributing instanton sector is fixed by chirality.
Therefore for an $n$-point function consisting of an arbitrary number of
vectors, $n_+$ positive chirality bilinears $(\bar\Psi (y)P_+ \Psi (x))$ and
$n_-$ negative chirality bilinears only the sector $k=n_+ -n_-$ contributes.

General $n$-point functions always may be decomposed into a sum of $n$-point
functions of definite chirality. In fact, this may be used to compute all
$n$-point functions from the trivial vacuum sector via clustering. E.g. for
the four-point function we find ($S_\pm (y,x)\equiv \bar\Psi (y)P_\pm \Psi
(x)$)
\begin{displaymath}
\langle TS_+ (y_2 ,y_1)S_- (x_2 ,x_1)\rangle \equiv \langle T\bar\Psi (y_2)P_+
\Psi (y_1)\bar\Psi (x_2)P_- \Psi (x_1)\rangle^{k=0} =
\end{displaymath}
\beq
\frac{(x_1 -y_2)^\mu}{2\pi (x_1 -y_2)^2}\frac{(x_2 -y_1)_\mu}{2\pi (x_2
-y_1)^2} \cdot e^{2{\rm\bf G}(0)+{\rm\bf G}(x_1 -x_2)+{\rm\bf G}(y_1 -y_2)-
\sum_{i,j}{\rm\bf G}(x_i -y_j)}.
\eeq
Performing now the limit $y_j \ra y_j +a$, $a\ra\infty$, $y_1 -y_2$ fixed, we find
that the free fermion propagators just cancel against the large range part of
$\exp \sum {\rm\bf G}(x_i -y_j)$ (the massless propagator $D_0$; the massive
one, $D_\mu$, vanishes exponentially). Comparing with the
expression (7) for the two-point function we therefore find
\beq
\lim_{a\to\infty}\langle TS_+ (y_2 +a,y_1 +a)S_- (x_2 ,x_1)\rangle^{k=0}=
\langle TS_+ (y_2 ,y_1)\rangle^{k=1}\langle TS_- (x_2 ,x_1)\rangle^{k=-1}
\eeq
which is the cluster property.

Generally, any VEV can be computed from the trivial sector by multiplying the operators
of given chiralities by their opposite chirality counterparts and by the use
of the cluster property.

\input psbox.tex

\section{Schwinger model perturbation theory}

In our perturbative calculations we will always use the exact photon
propagator, which consists of a sum of all possible vacuum polarization
insertions, 
  
$$\psboxscaled{800}{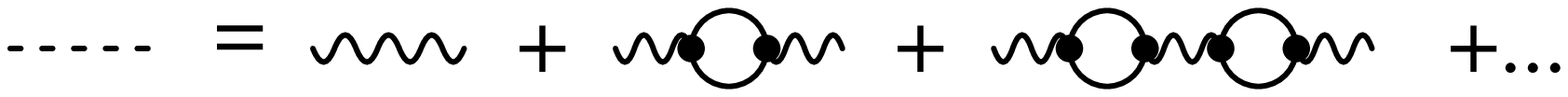}$$

and all other possible contributions vanish in the Schwinger model (for
details see e.g. \cite {Fr1}, \cite{BK1}, \cite{ABH}, \cite{Diss}).
In momentum space it is 
\beq
D^{\mu\nu}(p)=(g^{\mu\nu}-\frac{p^\mu p^\nu}{p^2})\frac{1}{p^2 +\mu^2}=-
\epsilon^{\mu\lambda}p_\lambda \epsilon^{\nu\rho}p_\rho\frac{1}{p^2
(p^2 +\mu^2)}
\eeq
as can be easily computed from the vacuum functional (1) or within
perturbation theory (\cite{ABH}). The propagator looks like being in "Landau
gauge", however this is the unique and gauge invariant result. When inserted
into a Feynman diagram, this propagator gives rise to
\beq
-e\gam\epsilon^{\mu\lambda}p_\lambda\ldots\frac{1}{p^2 (p^2 +\mu^2)}\ldots
e\gan \epsilon^{\lambda\rho}p_\rho =-\psla\gaf\ldots\psla\gaf\tilde{\rm\bf
G}(p^2), 
\eeq
$\tilde{\rm\bf G}(p^2)$ being the Fourier transform of ${\rm\bf G}(x)$, eq.
(4). 

In actual computations, when an exact photon line begins and ends at the same
fermion line, there is always an even number of $\gamma$ matrices between the
two $\gaf$, therefore we may omit them in the computations (in more general
cases we will reinsert them when it is necessary).

Further we will frequently use the identity
\beq
\frac{1}{\psla}\ksla\frac{1}{\psla +\ksla}=\frac{1}{\psla}-\frac{1}{\psla
+\ksla}.
\eeq
Let us take the fermionic two-point function as a first example. Up to second
order in the exact photon propagator we have

$$\psboxscaled{900}{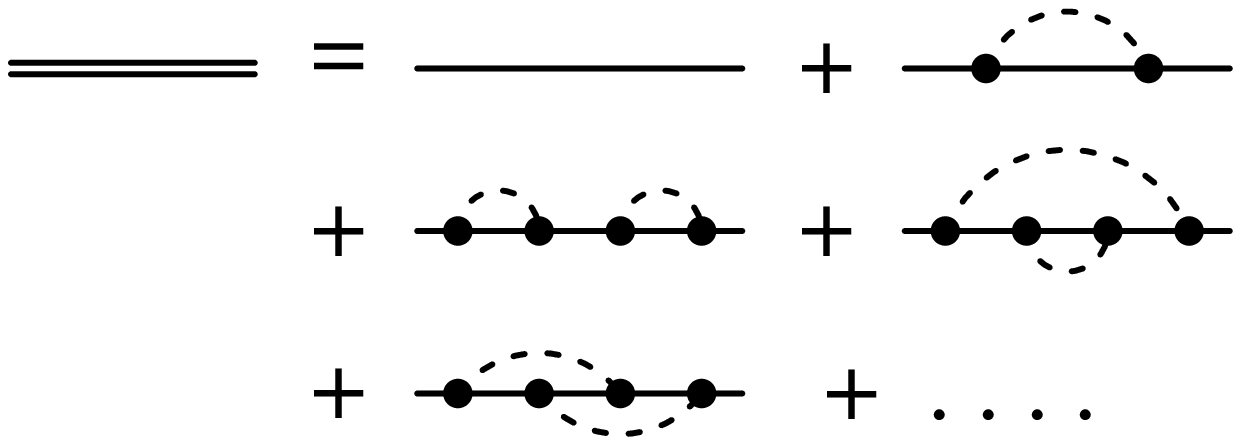}$$

For the first order graph

$$\psboxscaled{700}{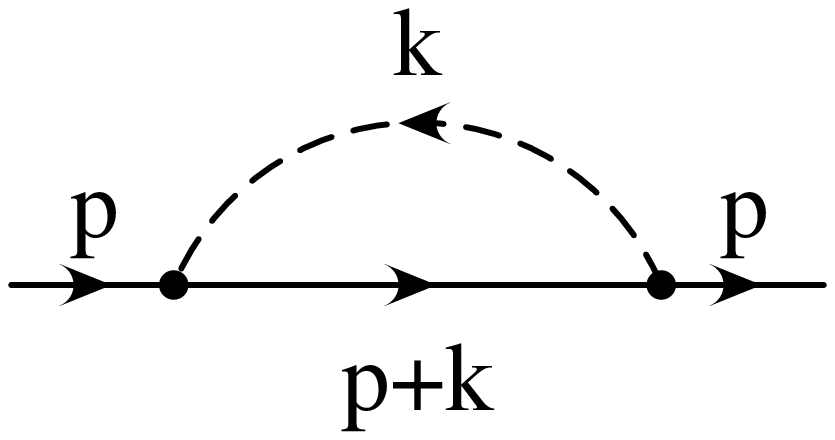}$$

we obtain, using (11) and (12)
\begin{displaymath}
-\int\frac{dk}{(2\pi)^2}{\rm\bf
G}(k)\frac{1}{\psla}\ksla\frac{1}{\psla +\ksla}\ksla\frac{1}{\psla}=
\end{displaymath}
\beq
-\int\frac{dk}{(2\pi)^2}{\rm\bf G}(k)\frac{1}{\psla}\ksla\frac{1}{\psla}+
\int\frac{dk}{(2\pi)^2}{\rm\bf G}(k)\frac{1}{\psla
+\ksla}\ksla\frac{1}{\psla} =
\eeq

$$\psbox{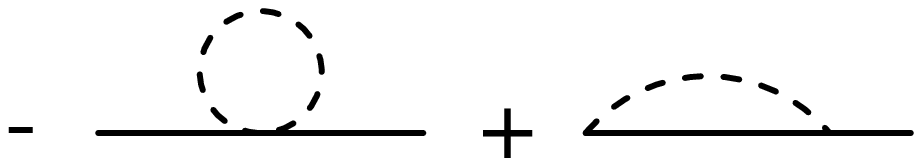}$$

and the first term vanishes. The final result is
\beq
\int\frac{dk}{(2\pi)^2}{\rm\bf G}(k)\frac{1}{\psla} - 
\int\frac{dk}{(2\pi)^2}{\rm\bf G}(k)\frac{1}{\psla +\ksla} =
\eeq

$$\psbox{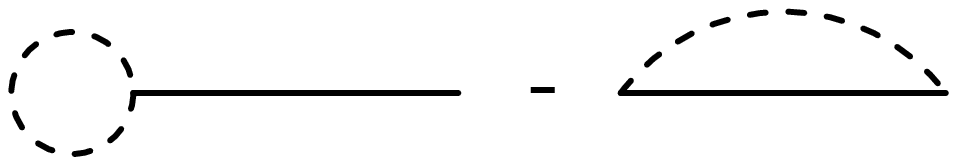}$$

which, after Fourier transformation, obviously is the first
order (in {\bf G}) expansion of the exact solution (6).

From this simple example we derive the following graphical rules: starting at
any vertex, you obtain two graphs by leaving out the inner or outer fermion
propagator (the latter one with a minus sign). Photon lines get a dot at that
vertex that has been changed (however photon lines running to the very begin
or end always have been changed, so it is not necessary there). Closed loops
vanish when only one of the two vertices has been changed. Otherwise they
decouple from all momenta and may be drawn to the begin or end of the fermion
line. Whenever a changed vertex coincides with an unchanged one, the unchanged
one cannot be removed further.

By applying these rules, we may reduce each individual graph to a final form.
Then we have to sum all graphs of a given order, and only the sum will lead to
the exact solution.

Applying these rules to the second order graphs we find

$$\psboxscaled{800}{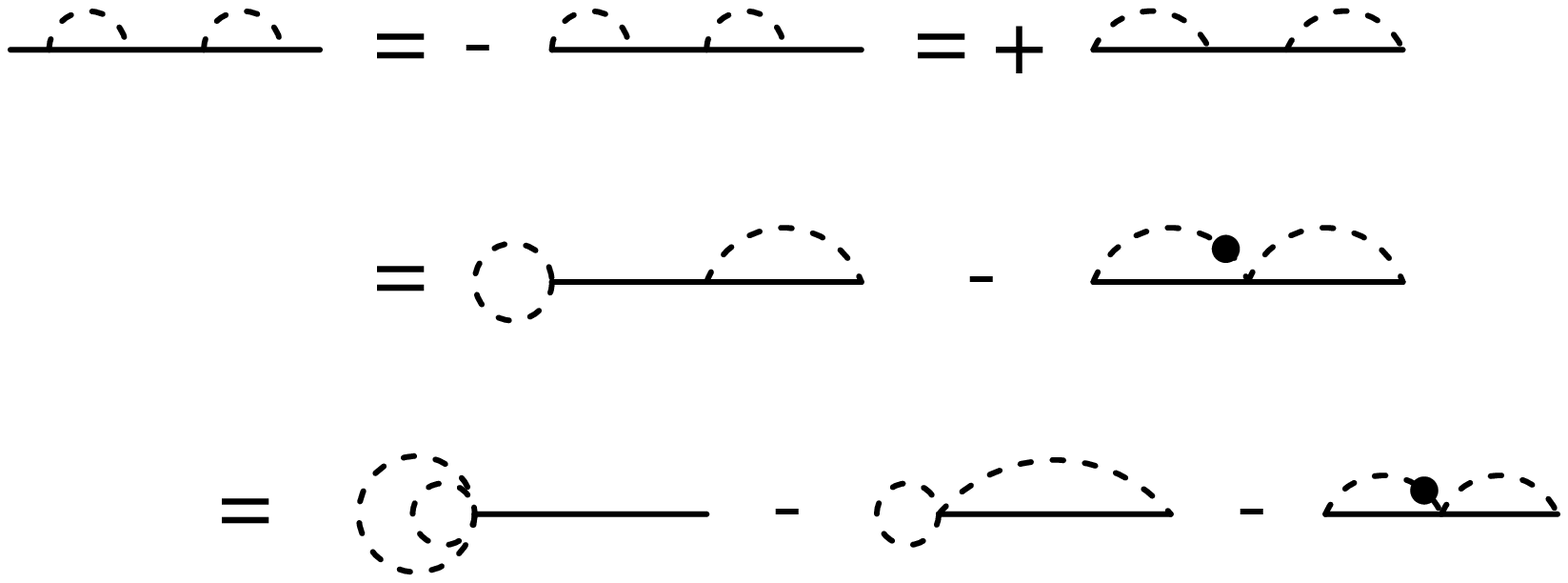}$$

$$\psboxscaled{800}{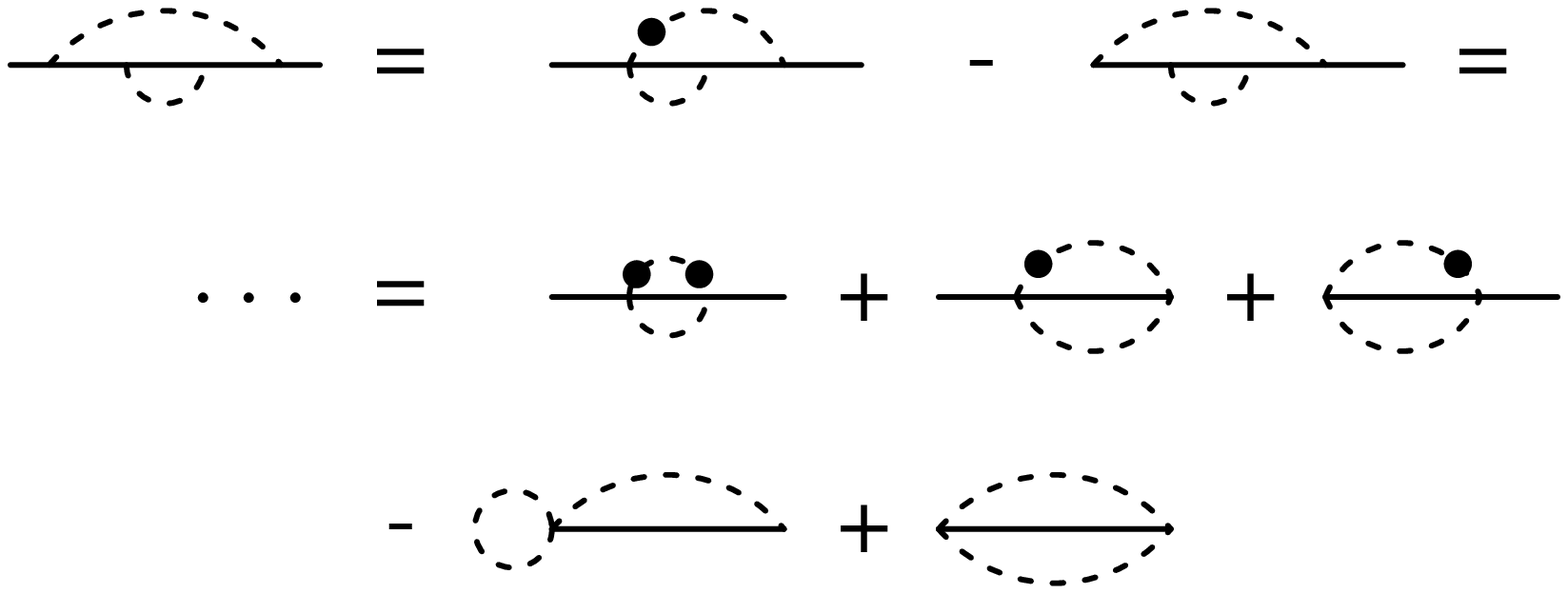}$$

$$\psboxscaled{800}{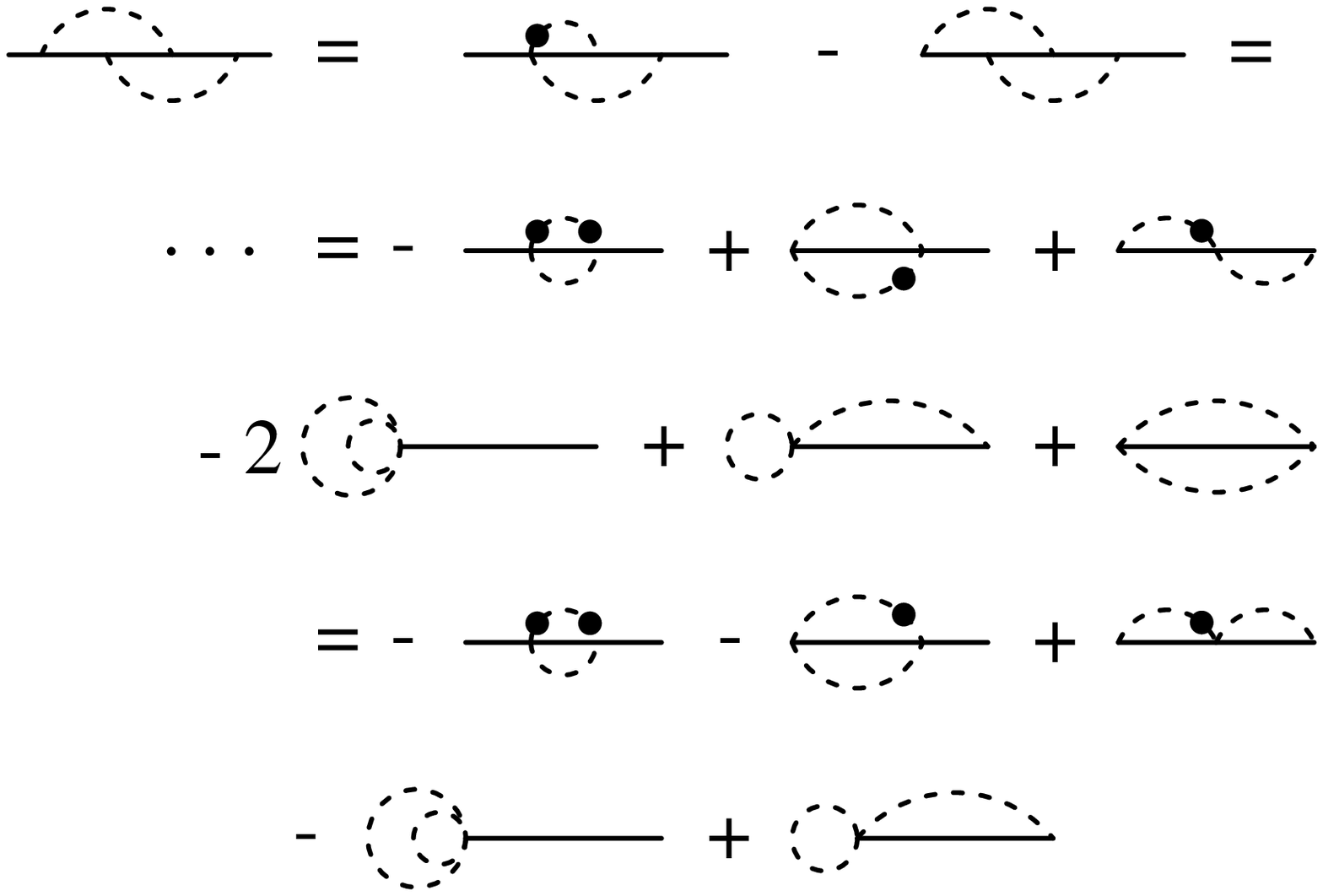}$$

where we used

$$\psboxscaled{800}{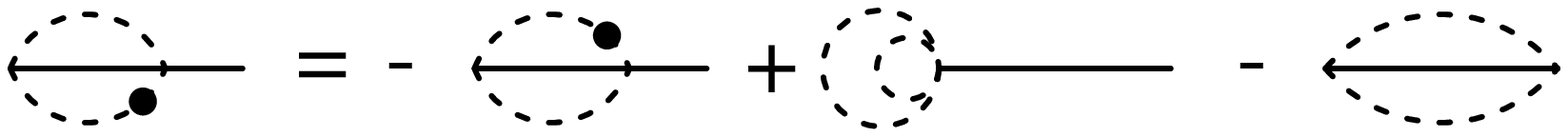}$$

\begin{displaymath}
\int d\mu (k,l)\frac{1}{\psla +\ksla +\lsla}\ksla\frac{1}{\psla}= -\int 
d\mu (k,l)\frac{1}{\psla +\ksla +\lsla}\lsla\frac{1}{\psla}+
\end{displaymath}
\beq
+\int d\mu (k,l)[\frac{1}{\psla}-\frac{1}{\psla +\ksla +\lsla}]
\eeq
in the last step (the symmetric integration measure $d\mu (k,l)$ we display in
a moment). For the sum of all three graph we get

$$\psboxscaled{800}{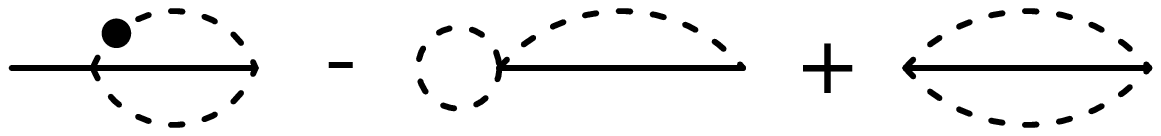}$$

In a next step we have to use the symmetry with respect to the two momentum
integrations, 

$$\psboxscaled{700}{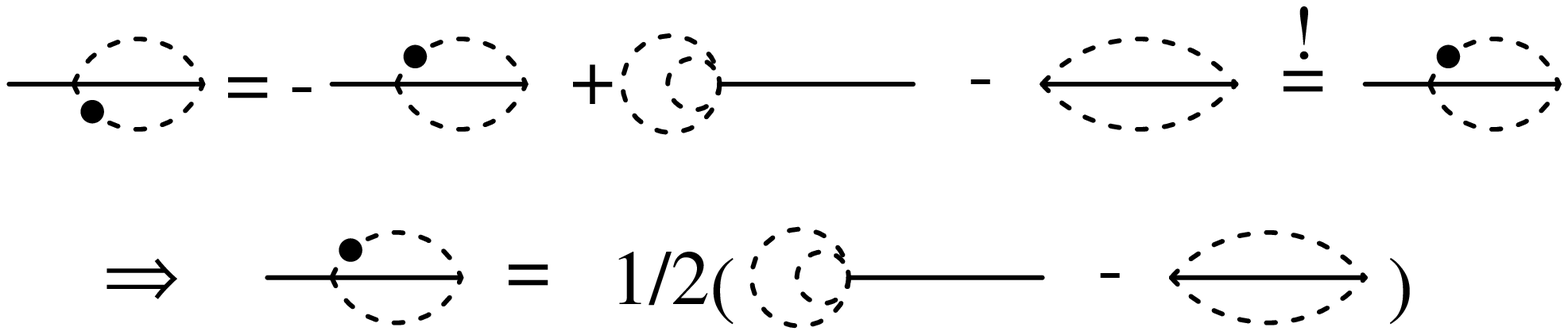}$$

to obtain the final result

$$\psboxscaled{800}{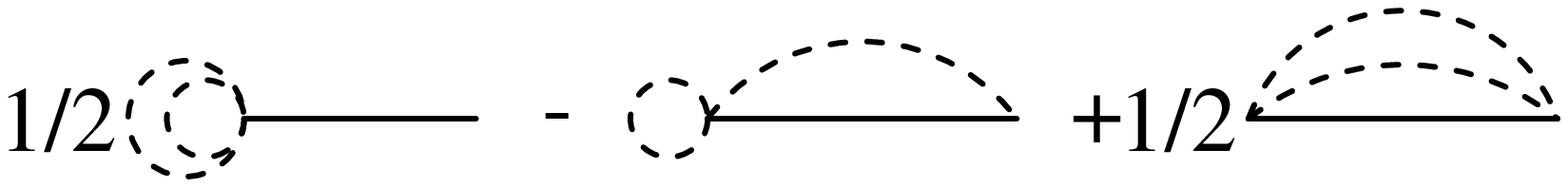}$$

\beq
\int d\mu (k,l)[\frac{1}{2}\frac{1}{\psla}-\frac{1}{\psla +\ksla}+frac{1}{2}
\frac{1}{\psla +\ksla +\lsla}],
\eeq
\begin{displaymath}
d\mu (k,l)=\frac{dk}{(2\pi)^2}\frac{dl}{(2\pi)^2}\tilde{\rm\bf G}(k)
\tilde{\rm\bf G}(l).
\end{displaymath}
This result, after a 
Fourier transformation, obviously is the second order (in {\bf G}) of the
exact solution (6).

So the general way of obtaining the exact solution is: reduce all graphs of a
given order according to the graphical rules, rearrange some dotted graphs
(like in (15)) in order to cancel them, and when some dotted graphs are left,
rewrite them by using symmetric momentum integration. This recipe remains the
same for higher $n$-point functions. Only when translating back into formulae
you have to insert one $\gaf$ matrix at the beginning of each fermion line
that carries an {\em odd} number of vertices.

E.g. for the exchange part of the four-point function 

$$\psboxscaled{700}{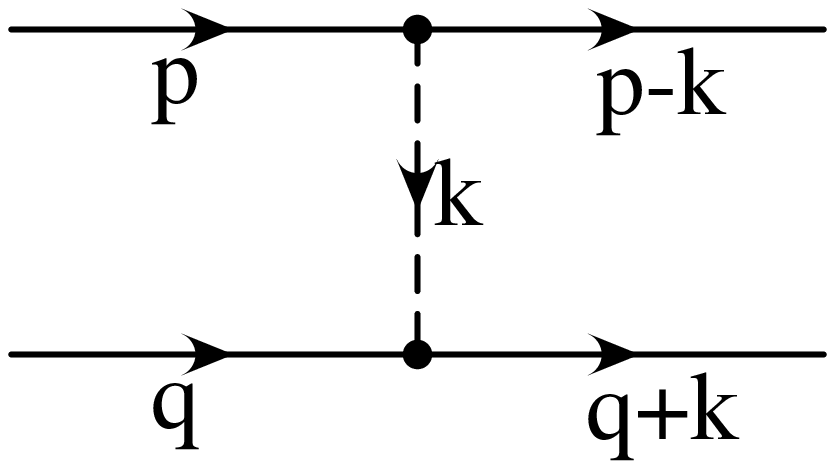}$$

one obtains

$$\psboxscaled{900}{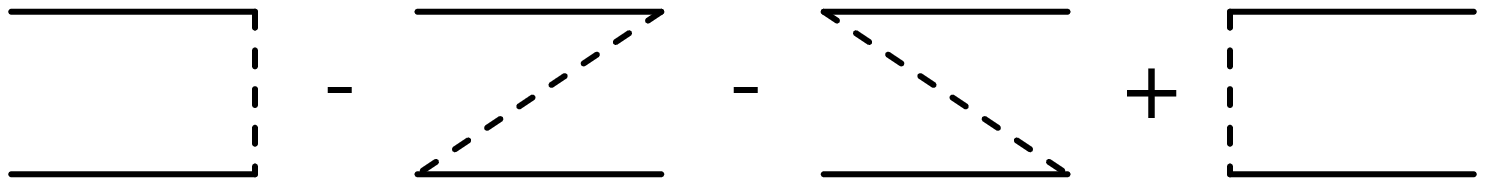}$$

for the first order and

$$\psboxscaled{700}{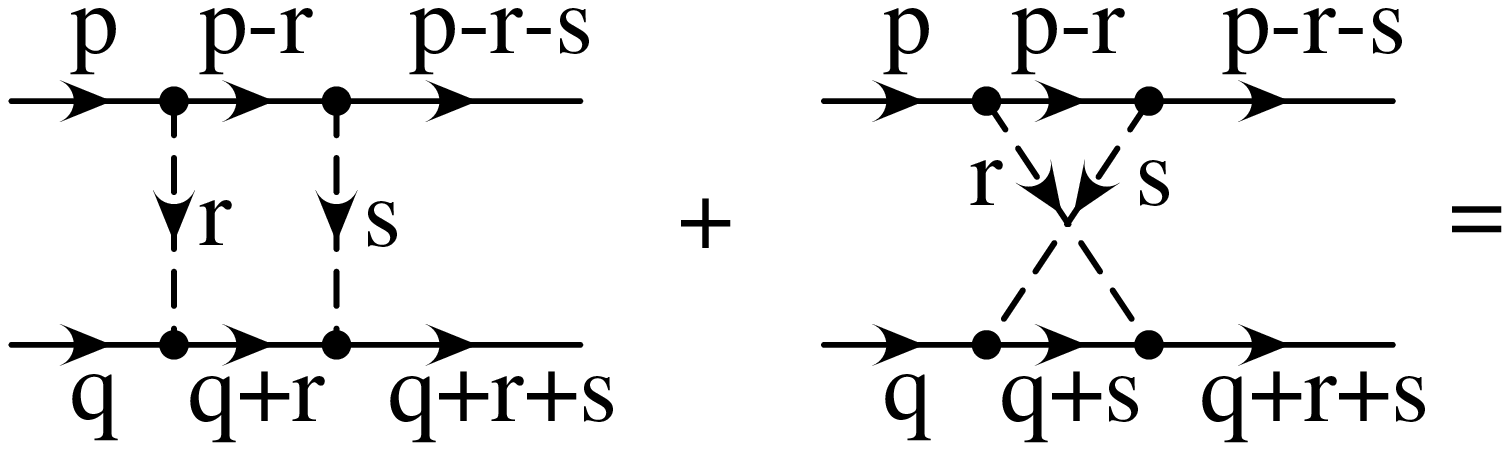}$$

$$\psboxscaled{800}{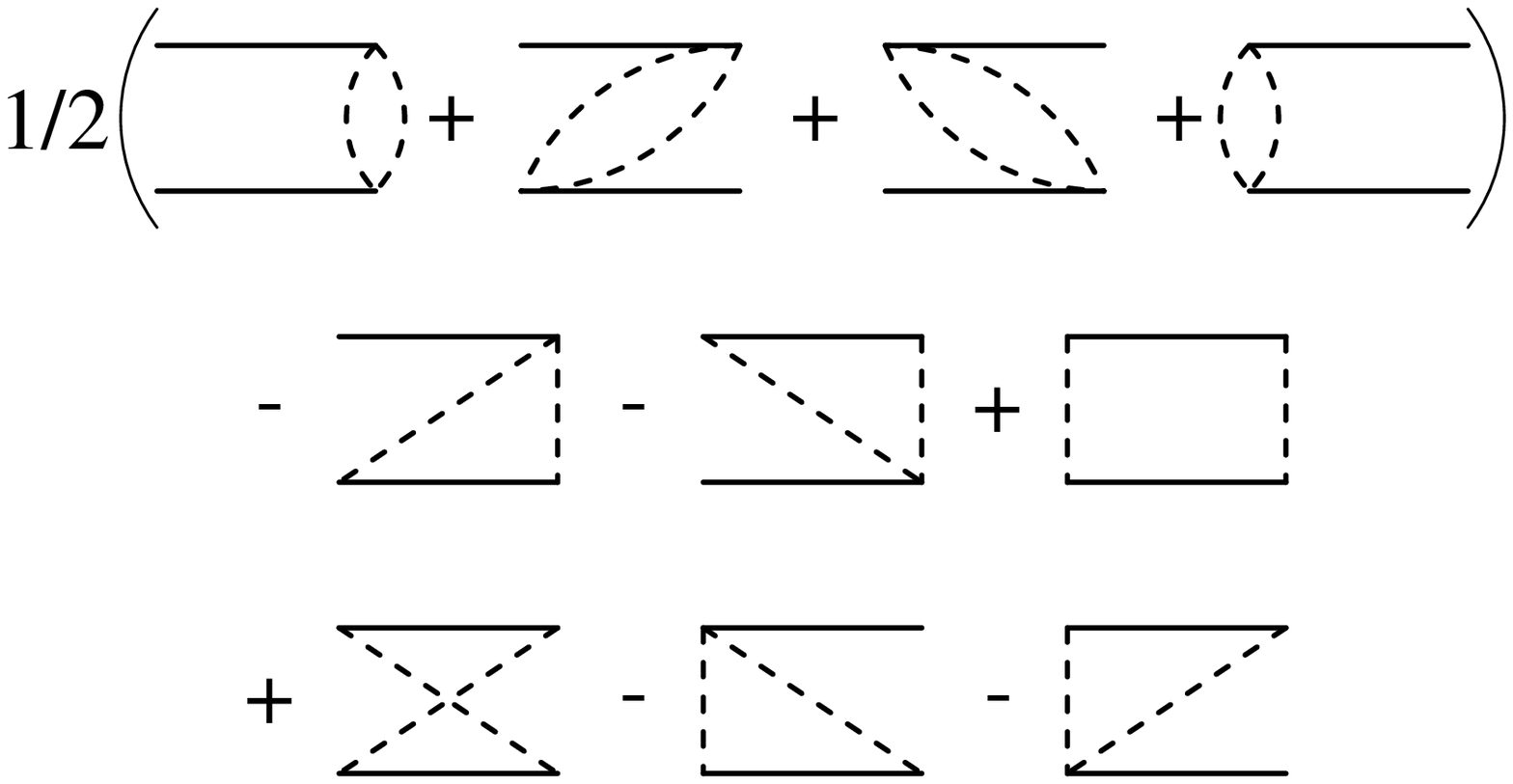}$$

in second order. These results coincide with the first and second order
expansions of the exact result (8).

So we found systematic rules to derive the exact solution order by order, but
the question remains if there is a way to derive the whole exact solution at
once from perturbation theory; and indeed there is. More precisely, for any
$n$-point function we may derive an exact integral equation from perturbation
theory. This we discuss in the next section.

\section{Integral equations for $n$-point functions}

Again, we start with the two-point function (for the two-point function an
analogous computation in light cone gauge has been done in \cite{Sta}). Let us
look at a graph of arbitrary order where the first photon runs from vertex 1
to some vertex $i$:

$$\psboxscaled{600}{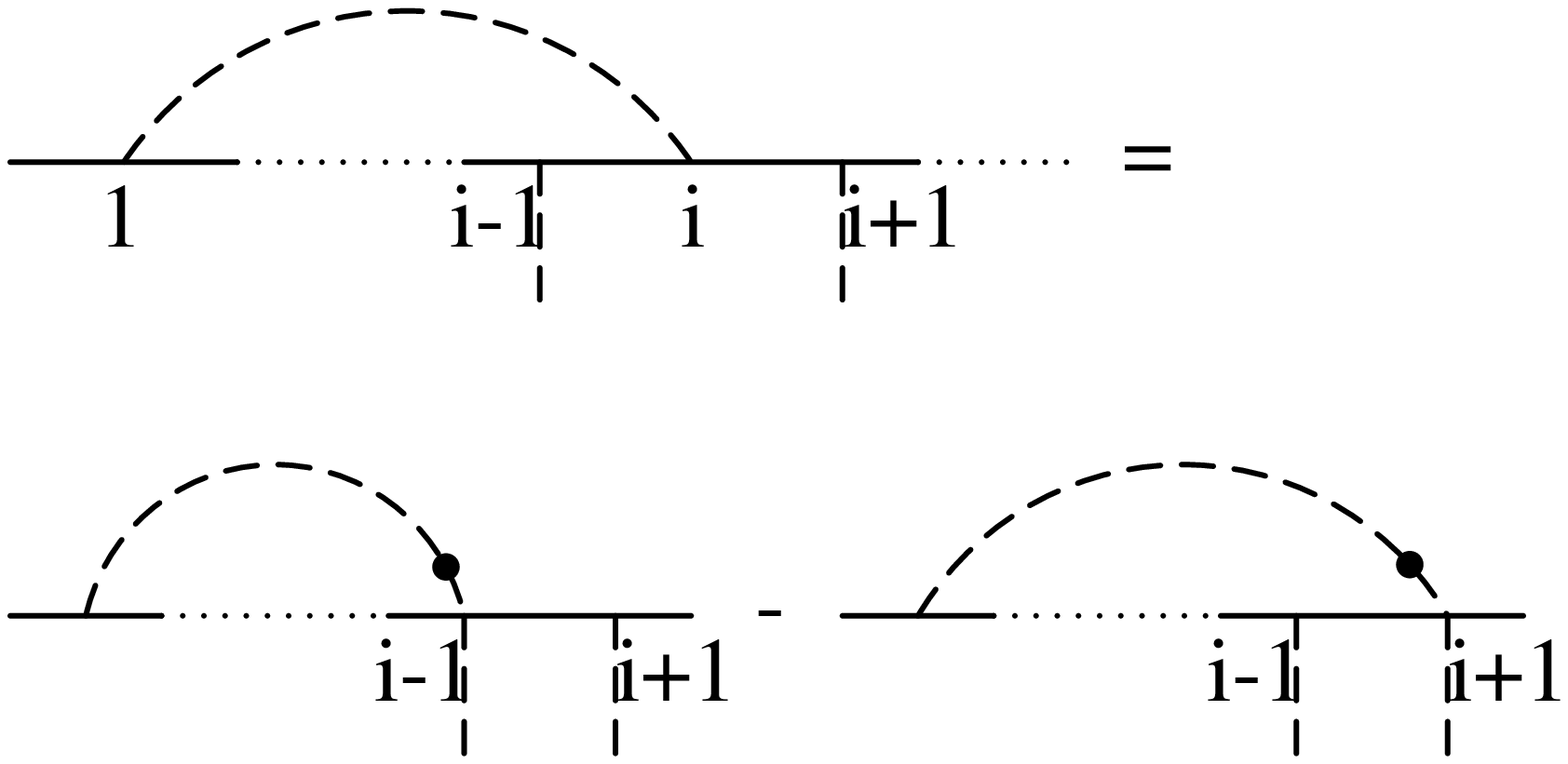}$$

whereas when running from 1 to $(i-1),(i+1)$ (all other lines being unchanged) we
find 

$$\psboxscaled{600}{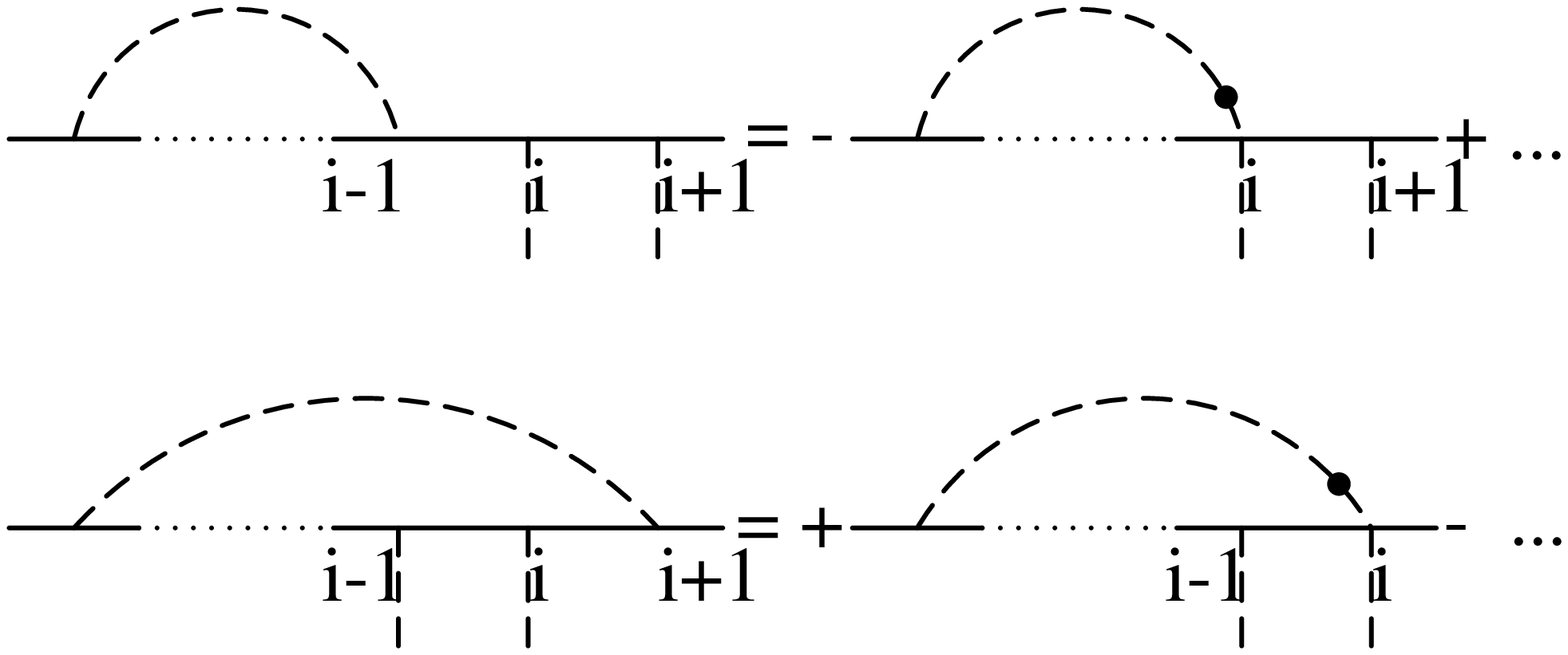}$$

and the diagram with $1\ldots i$ is cancelled by contributions from $1\ldots
(i-1)$ and $1\ldots (i+1)$ (the diagrams that cancel are indeed identical although
we indexed the vertices differently). Now, summing all graphs of a given order, 
for an arbitrary $1\ldots i$ diagram
there exist the corresponding $1\ldots (i-1)$ and $1\ldots (i+1)$ diagrams,
therefore nearly all diagrams cancel against each other. The exceptions are
the $1\ldots 2$ diagram, when vertex 2 is removed towards vertex 1 (this
however leads to a closed loop that is zero according to our rules), and the
$1\ldots 2n$ diagram (for  diagrams of order $n$).  
Here the vertex $2n$ has to be removed towards the very end of the graph.
Because this cancellation argument is true for arbitrary distributions of the
remaining photon lines, we get a recursion formula for the two-point function
of $n$th order:

$$\psboxscaled{700}{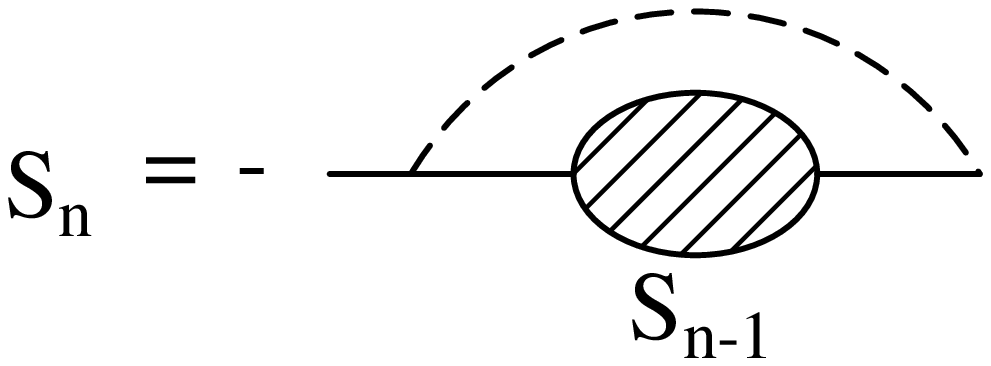}$$

\begin{displaymath}
S_n (p)=-\int\frac{dk}{(2\pi)^2}\tilde{\rm\bf G}(k)\frac{1}{\psla}\ksla
S_{n-1}(p-k)
\end{displaymath}
\beq
\psla S_n (p)=-\int\frac{dk}{(2\pi)^2}\tilde{\rm\bf G}(k)\ksla S_{n-1}(p-k)
\eeq
leading to the integral equation (with the zeroth order as inhomogenous part)
\beq
\psla S(p)=1-\int\frac{dk}{(2\pi)^2}\tilde{\rm\bf G}(k)\ksla S(p-k).
\eeq
It is very easy to see that the exact solution (6) solves this equation:
\begin{displaymath}
S(x)=G_0 (x)e^{{\rm\bf G}(0)-{\rm\bf G}(x)}
\end{displaymath}
\beq
\dsla S(x)=i\delta (x)-(\dsla {\rm\bf G}(x))S(x)
\eeq
which is the Fourier transform of the above equation. In fact, eq. (18)
is just the momentum space version of the Dyson-Schwinger equation for the
two-point function (for a general discussion of the Dyson-Schwinger equations
for the Schwinger model in $x$ space see \cite{DSEQ}; there the discussion is
for all instanton sectors).

The integral equation for the four-point function may be derived in a similar
fashion. Arguing in the same way like before we may derive

$$\psboxscaled{600}{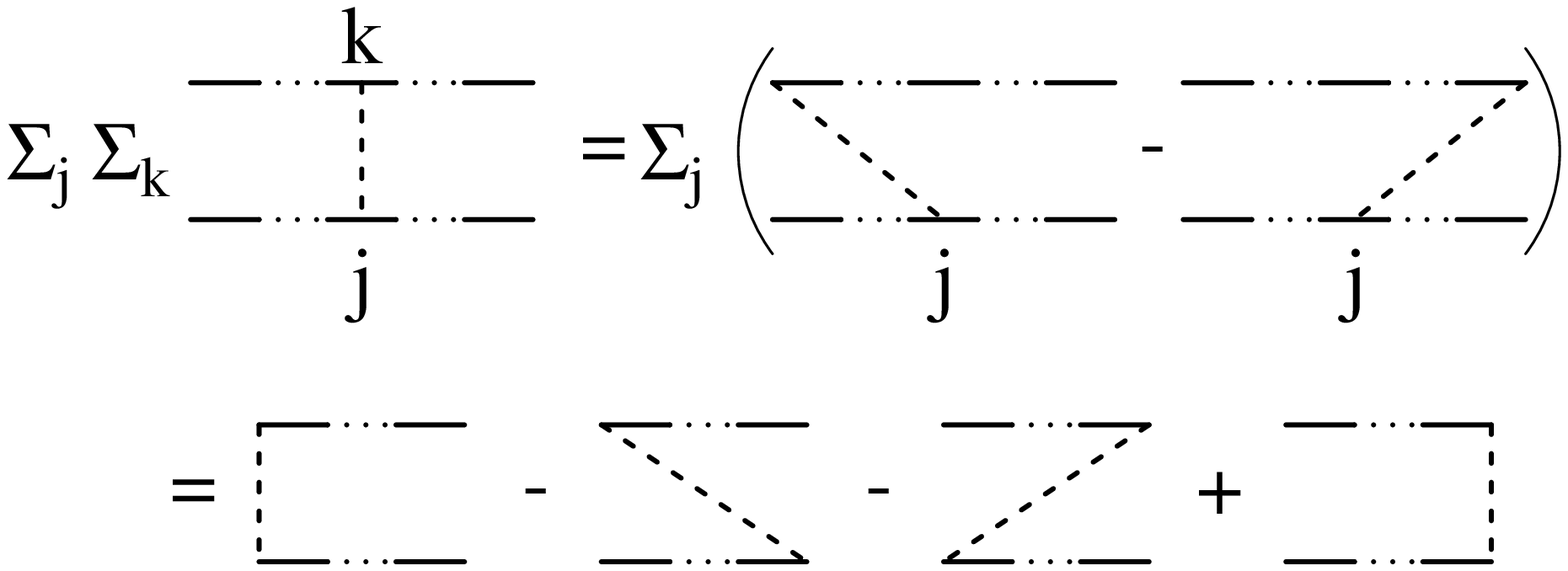}$$

or as a recursion formula

$$\psboxscaled{700}{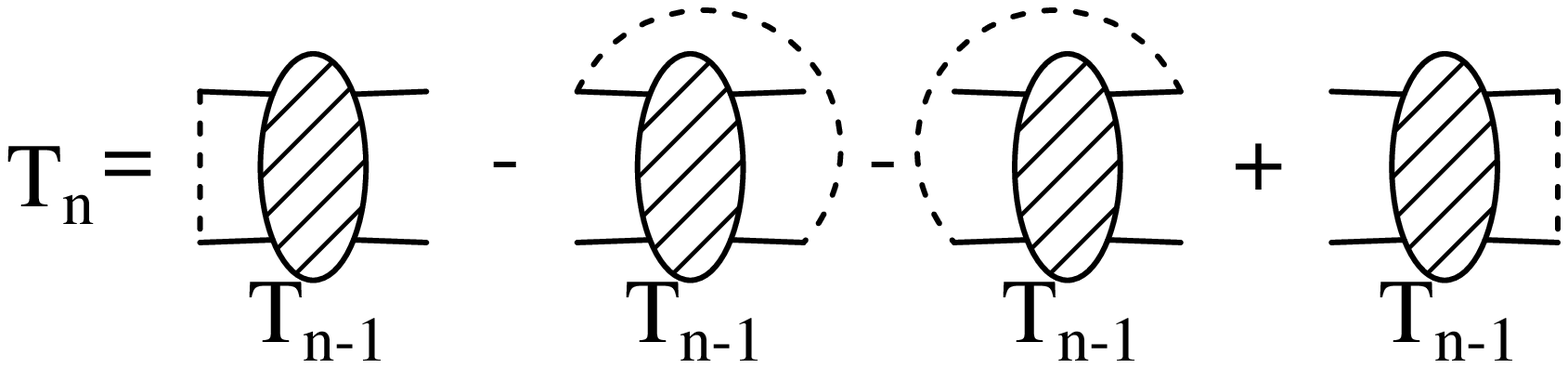}$$

\begin{displaymath}
T_n^{\alpha\beta\gamma\delta}(p,q,p-k,q+k)=\int\frac{dk_2}{(2\pi)^2}dk_1
\delta (k-k_1 -k_2)\tilde{\rm\bf G}(k_2)\gaf^{\alpha\alpha'}\gaf^{\beta\beta'}
\cdot
\end{displaymath}
\begin{displaymath}
\cdot [T_{n-1}^{\alpha'\beta'\gamma\delta}(p-k_2 ,q+k_2 ,p-k,q+k)-
T_{n-1}^{\alpha'\beta'\gamma\delta}(p-k_2 ,q,p-k,q+k_1)-
\end{displaymath}
\beq
T_{n-1}^{\alpha'\beta'\gamma\delta}(p,q+k_2 ,p-k_1 ,q+k)+
T_{n-1}^{\alpha'\beta'\gamma\delta}(p,q,p-k_1 ,q+k_1)]
\eeq
where however this operation only rises the number of {\em exchanging}
photons. The "self energy" part is, of course, computed like for the two-point
function, eq. (17). 

In order to derive an equation analogous to (18), we again multiply by
$\psla$, amputating thereby the upper incoming fermion line. As a consequence,
the first upper vertex is fixed, and we obtain (including now the self
energy):

$$\psboxscaled{700}{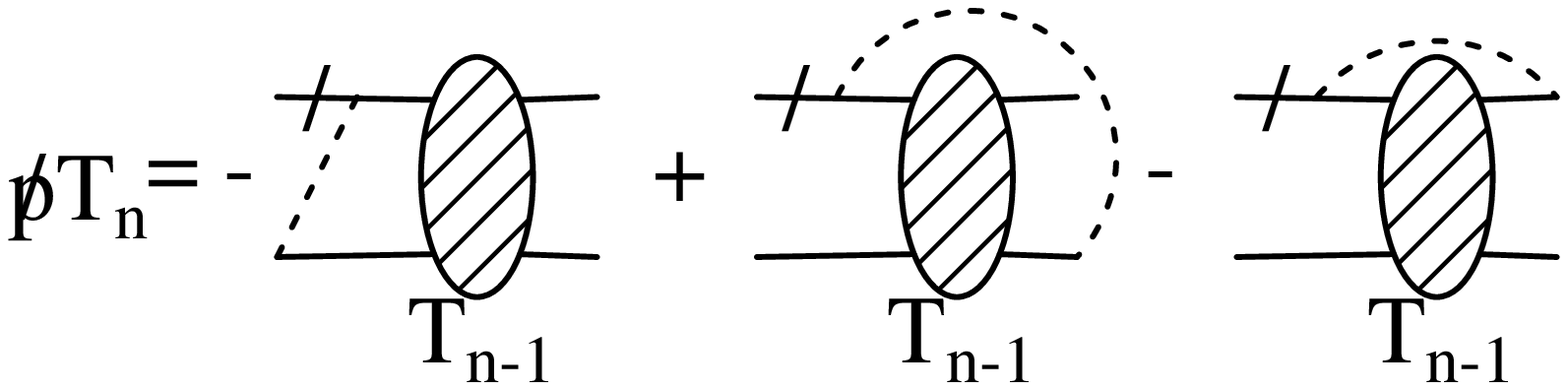}$$

(the slash indicates amputation) and therefore the equation (in symbolic
notation)
\bdi
\psla T(p,q,p-k,q+k)={\bf 1}\otimes S(q)\delta (k)+
\edi
\bdi
(\gam\gaf\otimes\gaf)\times\int\frac{dk_2}{(2\pi)^2}dk_1 \delta (k-k_1
-k_2) k_2^\mu \tilde{\rm\bf G}(k_2) \cdot
\edi
\bdi
\cdot [-T(p-k_2 ,q+k_2 ,p-k,q+k)+T(p-k_2 ,q,p-k,q+k_1 )]-
\edi
\beq
(\gam\otimes {\bf 1})\times\int\frac{dk_2}{(2\pi)^2}dk_1 \delta (k_1 -k-k_2)
k_2^\mu \tilde{\rm\bf G}(k_2) T(p-k_2 ,q,p-k_1 ,q+k)
\eeq
(for the inhomogenous part we used our knowledge of the two-point function).
This is the momentum space version of the Dyson-Schwinger equation for the
four-point function, see \cite{DSEQ}.

The generalization to higher $n$-point functions is straight forward.

\section{Summary}

We have given a systematic procedure of finding the exact $n$-point functions
from a resummed perturbation theory order by order. We have explicitly
demonstrated the procedure for the fermionic two- and four-point functions,
but generalizations are straight forward. Further we have derived integral
equations for the exact $n$-point functions from perturbative recursion
relations. By solving these equations the $n$-point functions in principle can
be found, which corresponds to a complete summation of the perturbation
series. By the way, these equations are just the Dyson-Schwinger equations (in
momentum space) which the exact $n$-point functions have to obey.

By these methods, and having in mind that the trivial instanton sector
suffices to obtain the complete solution of the Schwinger model, all features
of the Schwinger model in principle can be treated by perturbative methods.

\section*{Acknowledgements}
The author thanks H. Leutwyler and A. V. Smilga for very helpful discussions 
and the members of
the Institute of Theoretical Physics at Bern University, where this work was
done, for their hospitality.

This work was supported by a research stipendium of the University of Vienna.


\begin{thebibliography}{9999}
\bibitem{Sc1}
J. Schwinger, Phys. Rev. {\em 128} (1962) 2425
\bibitem{LS1}
J. Lowenstein, J. Swieca, Ann. Phys. {\em 68} (1971) 172
\bibitem{Jay}
C. Jayewardena, Helv. Phys. Acta {\em 61} (1988) 636
\bibitem{SW1}
I. Sachs, A. Wipf, Helv. Phys. Acta {\em 65} (1992) 653
\bibitem{DSEQ}
C. Adam, preprint UWThPh-1994-39; available on HEP-PH 95 01 273
\bibitem{Adam}
C. Adam, Z. Phys. {\em C63} (1994) 169
\bibitem{Diss}
C. Adam, thesis Universit\"at Wien 1993
\bibitem{ABH}
C. Adam, R. A. Bertlmann, P. Hofer, Riv. Nuovo Cim. {\em 16}, No
8 (1993)
\bibitem{Ja1}
R. Jackiw, in Treiman et al, "Current Algebras and Anomalies", World
Scientific, Singapore 1985
\bibitem{Sm1}
A. V. Smilga, Phys. Rev. {\em D46} (1992) 5598
\bibitem{Le1}
H. Leutwyler, Helv. Phys. Acta {\em 59} (1986) 201
\bibitem{KS1}
J. Kogut, P. Susskind, Phys. Rev. {\em D10} (1974) 3468, 
Phys. Rev. {\em D11} (1975) 1477, 3594
\bibitem{CKS}
A. Casher, J. Kogut, P. Susskind, Phys. Rev. {\em D10} (1974) 732
\bibitem{KS2}
J. Kogut, P. Susskind, Phys. Rev. {\em D11} (1975) 3594
\bibitem{BK1}
G. T. Bodwin, E. V. Kovacs, Phys. Rev. {\em D35} (1987) 3198
\bibitem{Fr1}
Y. Frishman, "Quark trapping in a model field theory", in: Lecture
Notes in Physics Nr. 32, Springer Verlag 1975
\bibitem{GS1}
R. E. Gamboa Saravi, M. A. Muschietti, F. A. Schaposnik, J. E.
Solomin, Ann. Phys. {\em 157} (1984) 360
\bibitem{DR1}
W. Dittrich, M. Reuter, "Selected Topics ...", Lecture Notes in
Physics {\em Vol.244}, Springer, Berlin 1986
\bibitem{Sta}
I. O. Stamatescu, T. T. Wu, Nucl. Phys {\em B143} (1978) 503
\end{thebibliography}
\end{document}